\documentclass[showpacs,showkeys,preprintnumbers, amsmath,amssymb]{revtex4}
\usepackage{graphicx}
\usepackage{epsfig}
\usepackage{bm}
\begin{document}

\title{Quasinormal Modes and Stability Criterion of Dilatonic Black Hole in $1+1$ and $4+1$ Dimensions}
\author{Ram\'on Becar$^{1}$, Samuel Lepe$^{2}$ and Joel Saavedra $^{2}$}
\address{$^{1}${\small Departamento de F\'{\i}sica, Universidad de Concepci\'on,}
 {\small Casilla 160 C, Concepci\'on, Chile}}
\address{ $^{2}${\small Instituto de F\'{\i}sica, Pontificia
Universidad Cat\'olica de Valpara\'{\i}so, Casilla 4950,
Valpara\'{\i}so.} }

\begin{abstract}
We study the stability of black holes that are solutions of the
dilaton gravity derived from string-theoretical models in two and
five dimensions against to scalar field perturbations, using the
Quasinormal Modes (QNMs) approach. In order to find the QNMs
corresponding to a black hole geometry, we consider perturbations
described by a massive scalar field non-minimally coupled to
gravity. We find that the QNM's frequencies turn out to be pure
imaginary leading to purely damped modes, that is in agreement
with the literature of dilatonic black holes. Our result exhibits
the unstable behavior of the considered geometry against the
scalar perturbations. We consider both the minimal coupling case,
i.e., for which the coupling parameter $\zeta$ vanishes, and the
case $\zeta=\frac{1}{4}$.
\end{abstract}


 \maketitle

\section{\label{sec:Int}Introduction}

Two-dimensional theories of gravity have recently attracted much
attention\cite{Robinson:2005pd,Myung:2000hk,Torii:1998gm} as
simple toy models that possess many features of gravities
in higher dimensions. They also have
black hole solutions which play important role in
revealing various aspects of spacetime geometry and quantization of
gravity, and are also related to string theory
\cite{Teo:1998kp,McGuigan:1991qp}.

On the other hand, there
is also a growing interest in five-dimensional
dilatonic black holes in the last few years, since it is believed that these
black holes can shed some light to the solution of the fundamental problem
of the microscopic origin of the Bekenstein-Hawking entropy. The
area-entropy relation $S_{BH} = A/4$ was obtained for a class of
five-dimensional extremal black holes in Type II string theory
using D-brane techniques \cite{Strominger:1996sh}, while in
Ref.\cite{Teo:1998kp} the U-duality that exists between
the five-dimensional black hole and the two-dimensional charged
black hole was exploited \cite{McGuigan:1991qp}
to microscopically compute the
entropy of the latter. For that reason, it is important to understand
the dynamics of matter fields and the metric perturbations in such
black hole backgrounds in order to find stable solutions. One of
the key issues worth of studying are so-called quasinormal modes (QNMs),
known as the ``ringing'' of black holes, that play an essential role in
the analysis of classical aspects of black holes physics.

In this work we are interested in the stability of the
1+1-dilatonic black hole using the QNMs' approach, quasinormal
modes associated with perturbations of different fields were
considered in different works \cite{Kokkotas:1999bd}, and for AdS
and dS space
\cite{Horowitz:1999jd,c2,c3,c4,c5,Chan:1999sc,Wang:2001tk,Konoplya:2003dd}.
Similar situation occurs in $2+1$ dimension
\cite{Chan:1996yk,c44,Crisostomo:2004hj}, and the acoustic black
holes \cite{c1, Lepe:2004kv,Saavedra:2005ug}. Quasinormal modes of
dilatonic black holes in $3+1$ dimensions can be see in
Refs.\cite{Ferrari:2000ep, Konoplya:2002ky,Fernando:2003wc}

 Determination of QNMs
for a specific geometry implies solving the field equations for
different types of perturbations (scalar, fermionic, vectorial,
etc.), with suitable boundary conditions that reflect the fact
that this geometry describes a black hole. Quasinormal modes for a
scalar classical perturbation of black holes are defined as the
solutions of the Klein-Gordon equation characterized by purely
ingoing waves at the horizon, $\Phi \sim e^{-i\omega (t+r)}$,
since at least a classically outgoing flux is not allowed at the
horizon. In addition, one has to impose boundary conditions on the
solutions in the asymptotic region (infinity), and for that it is
crucial to use the asymptotic geometry of the spacetime under
study. In the case of an asymptotically flat spacetime, the
condition we need to impose over the wave function is to have a
purely outgoing waves $\Phi \sim e^{-i\omega (t-r)}$ at the
infinity \cite{Horowitz:1999jd}. In general, the QNMs are given by
$\omega _{QNM}=\omega _{R}+i\omega _{I}$, where $\omega _{R}$ and
$\omega _{I}$ are the real and imaginary parts of the frequency
$\omega _{QNM}$, respectively. Therefore, the study of QNMs can be
implemented as one simple test for studying the stability of the
system. In this sense, any imaginary frequency with the wrong sign
would mean an exponentially growing mode, rather than a damping of
it.

In this work we analytically compute the QNMs of 1+1-dilatonic
black hole, in order to test stability of the
system. The organization of this article is as follows: In Sec.II
we specify the 1+1-dilatonic black hole. In Sec.III we determine
the QNMs and we establish a criterion for the stability of the system.
In Sec.IV
we study the problem of QNMs for the five-dimensional dilatonic
black hole. Finally, we finish with the conclusions in Sec.V.

\section{\label{sec:Dilatonic}$1+1$-Dilatonic Black Hole}

In order to have a gravity theory with dynamical degrees of
freedom in two-dimensional spacetime, we consider the gravity coupled to a
dilatonic field described by the action
\begin{equation}
S_{g}=\frac{1}{2\pi }\int d^{2}x\sqrt{-g}e^{-2\phi }\left( R+4(\nabla \phi
)^{2}+4\lambda ^{2}\right) .  \label{accion}
\end{equation}
It is worthwhile notocing that the two-dimensional critical string theory
\cite{Witten:1991yr} has been an inspiration of
many articles, since it is a simple toy model possessing black hole solutions which
can be a starting
point to solve the problems of Hawking radiation and the information
loss inside black holes
\cite{Rahaman:2006wm,Hong:2005rn,Nojiri:2000eb,Ghosh:1994wb}.

It was also proved some time ago, that the dilatonic black hole is a
solution of an exact conformal field theory, namely the WZW model
with gauge group SL(2,R)/U(1). This solution can be
derived by solving the two-dimensional beta function equations of
the string theory, that is effectively a two-dimensional
graviton-dilaton system. The equations of motion for
the graviton and dilaton are given by
\begin{eqnarray}
\beta _{\mu \nu }^{G} &=&R_{\mu \nu }+2\nabla \mu \nabla \nu \phi
=0,\label{betag} \\
\beta ^{\phi } &=&\Box \phi -2\left( \nabla
\phi \right) ^{2}+2\lambda ^{2}=0. \label{betaphi}
\end{eqnarray}
A general static
metric describing a black hole in this theory can be written as
\begin{equation}
ds^{2}=-f(r)d\tau ^{2}+\frac{dr^{2}}{f(r)},  \label{metrica1}
\end{equation}
where $f(r)=1-e^{-\phi }$ and $\phi =(r-r_{0})/r_{0}$. If we change
the coordinate as $x=\frac{r-r_{0}}{r_{0}},$ then
the function $f(r(x))=f(x)$ becomes $f(x)=1-e^{-x}$ and the horizon of the black hole is located at
$x=0$. This solution represents a well-known string-theoretic
black hole
\cite{Teo:1998kp,McGuigan:1991qp,Witten:1991yr,Frolov:2000jh}.

\section{\label{sec:Dilatonic1}Quasinormal Modes}

In order to study the QNMs, we consider a scalar field with
no-minimal coupling to gravity, propagating in the background of
the dilatonic black hole. This system is described by
the action \cite{Frolov:2000jh}
\begin{equation}
S[\varphi ]=-\frac{1}{2}\int d^{2}x\sqrt{-g}\left( (\nabla \varphi
)^{2}+\left( m^{2}+\zeta R\right) \varphi ^{2}\right) ,  \label{accion2}
\end{equation}
where $\zeta $ is a parameter from the no-minimal coupling. The field
equations read
\begin{equation}
\left( \Box -\mu ^{2}-\zeta R\right) \varphi =0,
\label{fieldequations}
\end{equation}
where $\mu =r_{0}m$. In terms of the coordinate $x$ and assuming a
solution in the form $\varphi =e^{-i\omega t}R(x)$, the radial
equation (\ref{fieldequations}) can be written as
\begin{equation}
f\partial _{x}^{2}R(x)+e^{-x}\partial _{x}R(x)-(\frac{\omega ^{2}}{f}-\mu
^{2}-\zeta e^{-x})R(x)=0.  \label{requation1}
\end{equation}
Next, we define a new variable, $z=1-e^{-x}$, so that the radial equation adopts
the form
\begin{equation}
z(1-z)\partial _{z}\left( z(1-z)\partial _{z}R(z)\right) +\left(
\omega ^{2}-z\mu ^{2}-\zeta ^{\prime }z(1-z)\right) R(z)=0,
\label{zrdez}
\end{equation}
where $\zeta ^{\prime }=\zeta /r_{0}^{2}$ is a new parameter. With the change
$R(z)=z^{\alpha }(1-z)^{\beta }F(z)$, the last equation reduces to
the hypergeometric differential equation for the function $F(z)$, that is,
\begin{equation}
z(1-z)F^{\prime \prime }(z)+(c-(a+b+1)z)F^{\prime }(z)-abF(z)=0.
\label{hyper}
\end{equation}
Here, the coefficients $a$, $b$ and $%
c$ are given through the relations
\begin{eqnarray}
c &=&2\alpha +1,  \nonumber\\
a+b &=&2(\alpha +\beta )+1,  \label{coeficientes}  \\
ab &=&(\alpha +\beta )(\alpha +\beta +1)-\zeta ^{\prime },
\nonumber
\end{eqnarray}
from where we obtain the expressions for the coefficients,
\begin{eqnarray}
a &=&\frac{1}{2}\left( 1+2\alpha +2\beta -\sqrt{1-4\zeta ^{\prime
}}\right) ,\label{a1}
\\
b &=&\frac{1}{2}\left( 1+2\alpha +2\beta +\sqrt{1-4\zeta ^{\prime
}}\right),\label{b1}
\end{eqnarray}
and for the exponents $\alpha$ and $\beta$,
\begin{eqnarray}
\alpha &=&\pm i\omega ,  \label{alhpabeta} \\
\beta &=&\pm \sqrt{\omega ^{2}-\mu ^{2}}.  \label{lhpabeta}
\end{eqnarray}
Without loss of generality, above we choose the negative signs.
It is well-known that the hypergeometric equation has
three regular singular points, at $z=0,$ $z=1$ and $z=\infty $, and
it has two independent solutions in the neighborhood of each point
\cite{Abra}. The solutions of the radial equation reads as follows,
\begin{equation}
F(z)=C_{1}F_{1}(a,b,c;z)+C_{2}z^{1-c}F_{1}(a-c+1,b-c+1,2-c;z),
\label{hyper2}
\end{equation}
where $F_{1}(a,b,c;z)$ is the hypergeometric function and $C_1$, $C_2$ are constants.
The solution for $R(z)$ is then
\begin{equation}
R(z) =C_{1}z^{-i\omega }(1-z)^{-i\sqrt{\omega ^{2}-\mu
^{2}}}F_{1}(a,b,c;z)+ C_{2}z^{i\omega }(1-z)^{-i\sqrt{\omega
^{2}-\mu ^{2}}}F_{1}(a-c+1,b-c+1,2-c;z).\label{rdez}
\end{equation}
Note that, when $c=1$, two solutions become linearly dependent and
the general solution represents a bound state. This point was
discussed in Ref. \cite{Frolov:2000jh}.

In the neighborhood of the horizon $(z=0)$, the function $R(z)$ behaves as
\begin{equation}
R(z)=C_{1}e^{-i\omega \ln z}+C_{2}e^{i\omega \ln z},  \label{rnearH}
\end{equation}
and the scalar field $\varphi $ can be written in the following way,
\begin{equation}
\varphi \sim C_{1}e^{-i\omega (t+\ln z)}+C_{2}e^{-i\omega (t-\ln z)}.
\label{phinearH}
\end{equation}
From the above expression it is easy to see that the first term corresponds to an
ingoing wave, while the second one represents an outgoing wave
in the black hole. For computing the QNMs, we have to impose that
there exist only ingoing waves at the horizon so that, in order to
satisfy this condition, we set $C_{2}=0.$ Then the radial solution
at the horizon is given by
\begin{equation}
R(z)=C_{1}z^{-i\omega }(1-z)^{-i\sqrt{\omega ^{2}-\mu ^{2}}}F_{1}(a,b,c;z).
\label{rdez2}
\end{equation}
In order to implement the boundary conditions at the infinity $(z=1)$, we
use the linear transformation $z\rightarrow 1-z$, and then we apply the Kummer's formula
\cite{Abra} for the hypergeometric function. We obtain
\begin{eqnarray}
R(z) &=&C_{1}z^{-i\omega }(1-z)^{-i\sqrt{\omega ^{2}-\mu
^{2}}}\frac{\Gamma (c)\Gamma (c-a-b)}{\Gamma (c-a)\Gamma
(c-b)}F_{1}(a,b,a+b-c+1;1-z)+ \nonumber \\
&&+C_{1}z^{-i\omega }(1-z)^{i\sqrt{\omega ^{2}-\mu
^{2}}}\frac{\Gamma (c)\Gamma (a+b-c)}{\Gamma (a)\Gamma
(b)}F_{1}(c-a,c-b,c-a-b+1;1-z) \label{rinfinito}.
\end{eqnarray}
The above solution near the infinity ($z=1$) takes on the form
\begin{equation}
R(z) =C_{1}(1-z)^{-i\sqrt{\omega ^{2}-\mu ^{2}}}\frac{\Gamma
(c)\Gamma (c-a-b)}{\Gamma (c-a)\Gamma
(c-b)}+C_{1}(1-z)^{i\sqrt{\omega ^{2}-\mu ^{2}}}\frac{\Gamma
(c)\Gamma (a+b-c)}{ \Gamma (a)\Gamma (b)}, \label{rinfinito3}
\end{equation}
while the solution for the scalar field near the infinity behaves as
\begin{equation}
\varphi  \sim C_{1}e^{-i\sqrt{\omega ^{2}-\mu ^{2}}(t+\ln (1-z))}
\frac{\Gamma (c)\Gamma (c-a-b)}{\Gamma (c-a)\Gamma
(c-b)}+C_1 e^{-i\sqrt{\omega ^{2}-\mu ^{2}}(t-\ln (1-z))}\frac{\Gamma
(c)\Gamma (a+b-c)}{\Gamma (a)\Gamma (b)}\label{rinfinito4}.
\end{equation}
In order to compute the QNMs, we also need to impose the boundary
conditions on the solution of the radial equation at infinity,
meaning that only purely outgoing waves are allowed there. Therefore, the
second term in the above expression should be zero, what is fulfilled
only at the poles of $\Gamma (a)$ or $\Gamma (b)$. Since the gamma
function $\Gamma(x)$ has the poles at $x=-n$ for $n=0,1,2,...$,
the wave function satisfies the considered boundary condition only upon the
following additional restriction,
\begin{eqnarray}
a &=&-n,  \label{condicion} \\
&&\text{or}  \nonumber \\
b &=&-n,  \label{condicion1}
\end{eqnarray}
where $n=0,1,2,...\,$. These conditions
determine the form of the quasinormal modes, that is, from Eqs.(\ref{a1}) and
(\ref{b1}), we find
\begin{equation}
\omega  =-\frac{i}{4}\left( 1-\sqrt{1-4\zeta ^{\prime
}}-\frac{(1+\sqrt{1-4\zeta ^{\prime }})\mu ^{2}}{n+n^{2}+\zeta
^{\prime }} + n \left(2-\frac{2\mu ^{2}}{n+n^{2}+\zeta
^{\prime}}\right) \right).\label{frecuencias}
\end{equation}
The expression (\ref{frecuencias}) for frequencies shows a possible instability of
the black hole under scalar perturbations, that could imply an
exponentially growing mode if the wrong sign of the pure imaginary
frequency had been chosen (positive). This issue is clarified in Figs.(1)
and (2).
\begin{figure}[th]
\includegraphics[width=5.0in,height=4.0in,angle=0,clip=true]{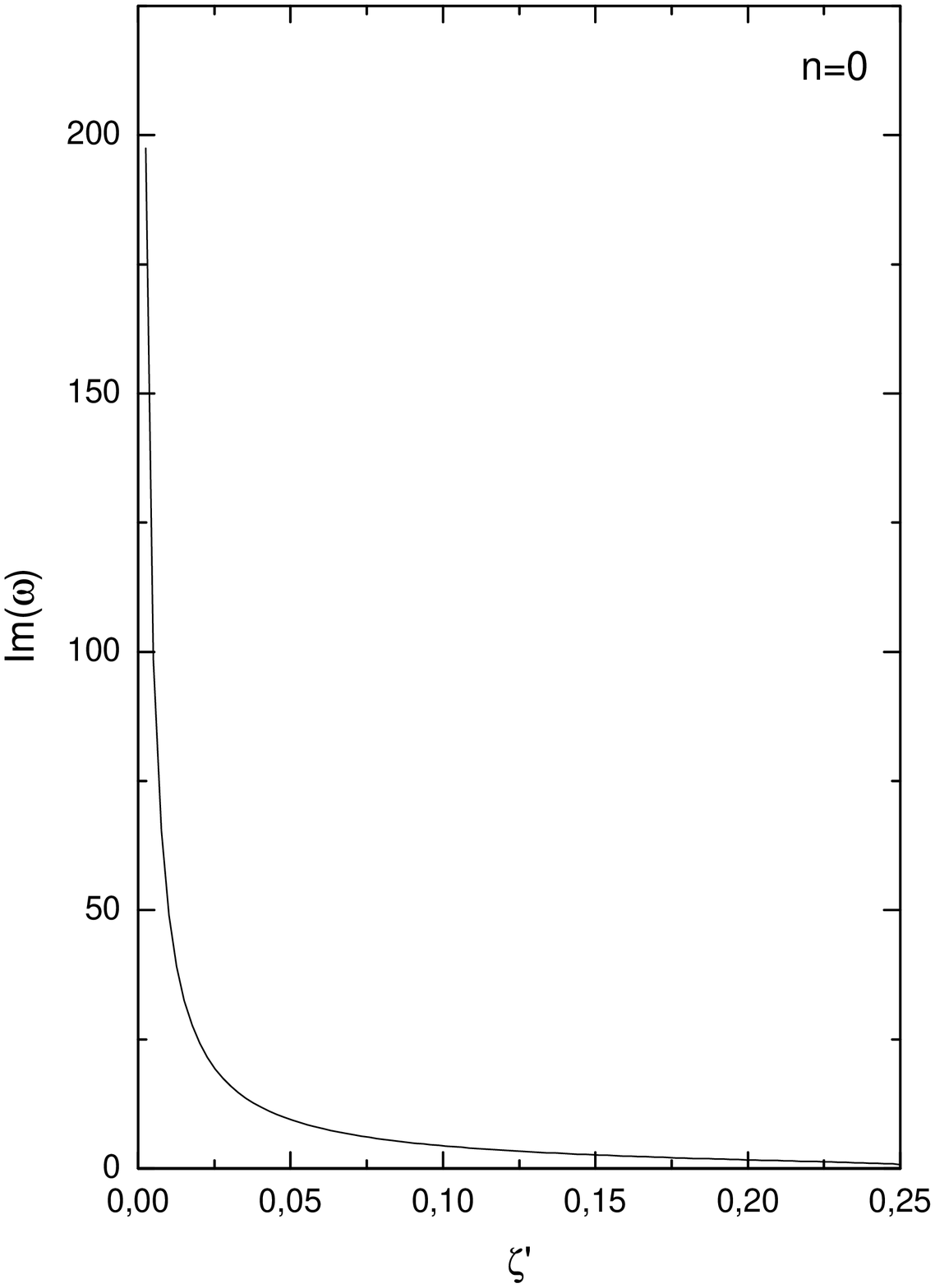}
\caption{The imaginary part of the QNM's frequency of the fundamental
mode as a function of the no-minimal coupling parameter. This plot
shows an unstable behavior of a scalar perturbation that excites the
fundamental mode. We have taken $\mu = 1$. } \label{fig1}
\end{figure}
Fig.(1) shows the instability arising in the fundamental modes for
scalar perturbations that excite these modes, in the range $0\leq
\zeta'\leq1/4$. Note that in this range of the no-minimal coupling
parameter the quasinormal modes are purely imaginary, as in the
2+1-dilatonic case \cite{Fernando:2003ai}. The plot in the figure
corresponds to the mass $\mu=1$. If we consider an arbitrary mass
for the scalar field, then the instability it is also present, and
depends on the values of $\mu$ with respect to $n$. This fact can
be explicitly shown for $\zeta ^{\prime }=0$ (minimal coupling),
when we obtain for the frequency
\begin{equation}
\omega =-i\frac{(n^{2}-\mu ^{2})}{2n}.  \label{wminimal}
\end{equation}
We see that the overtones $n<\mu $  guarantee the instability
under scalar perturbations in particular the fundamental mode as
is show in Fig. (1). A similar situation occurs in the conformal
case $\zeta ^{\prime }=1/4$, where
\begin{equation}
\omega =-i\frac{\left( \left( 1+2n\right) ^{2}-4\mu ^{2}\right)
}{4+8n}, \label{w1/4}
\end{equation}
\ \ if $n<\mu -1/2$. In summary the two dimensional dilatonic blak
hole show an unstable behavior against scalar perturbations this
results was shown in Ref.\cite{Azreg-Ainou:1999kb} where the
instability of 1+1 dilatonic black holes has been shown using
metric perturbations.
\begin{figure}[th]
\includegraphics[width=5.0in,height=5.0in,angle=0,clip=true]{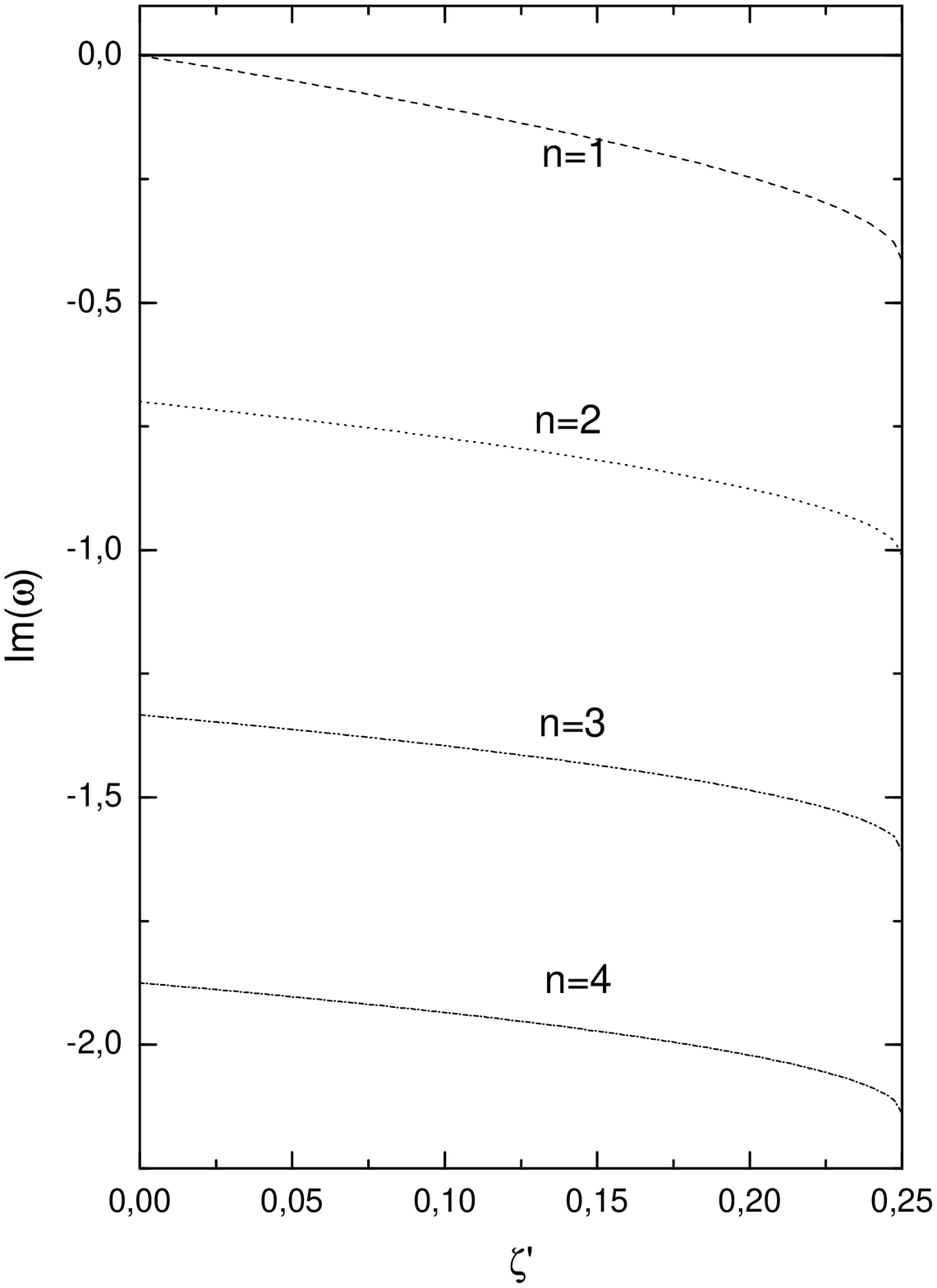}
\caption{The imaginary part of the QNM's frequency as a function
of the no-minimal coupling parameter, for several overtones. This
plot shows a stable behavior of scalar perturbations for all
overtones. We have taken $\mu = 1$. } \label{fig1}
\end{figure}
In the range of parameters $\zeta^{\prime}>1/4$, the frequency of
QNMs acquires a real part,
\begin{equation}
\omega =-\frac{\sqrt{4\zeta ^{\prime }-1}}{4}\left( 1+\frac{\mu ^{2}}{%
n+n^{2}+\zeta ^{\prime }}\right) -\frac{i}{4}\left( 1--\frac{\mu ^{2}}{%
n+n^{2}+\zeta ^{\prime }}+n\left( 2-\frac{2\mu ^{2}}{n+n^{2}+\zeta ^{\prime }%
}\right) \right) .  \label{frecuencias2}
\end{equation}
Figure (3) shows the behavior of both the real and imaginary parts
of QNMs. In this range, we observe that the black hole is stable
for all QNMs for $\zeta'>1$.

\begin{figure}[th]
\includegraphics[width=5.0in,height=6.0in,angle=0,clip=true]{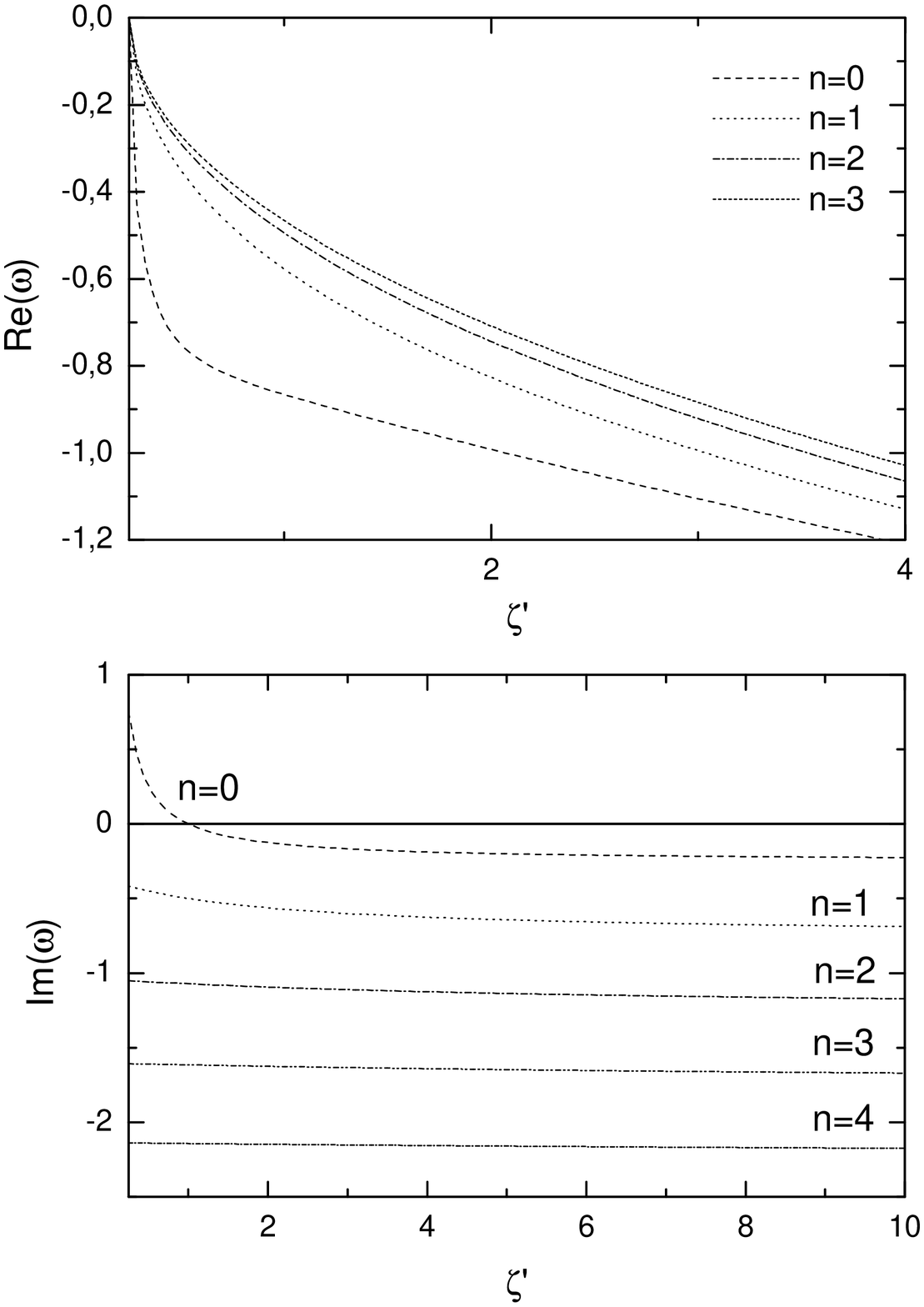}
\caption{The upper panel shows the real part of the QNM's frequency as
a function of the no-minimal coupling parameter for several
overtones, in case of the two-dimensional black hole. Note that,
for a high no-minimal parameter, the real part coalesce. The lower
panel shows the imaginary part of the QNM's frequency as a function of
the no-minimal coupling parameter, for several overtones. It demonstrates a
stable behavior of scalar perturbations for all overtones with
$\zeta'>1$. We have taken $\mu = 1$. } \label{fig1}
\end{figure}

Finally, note that the real part of the QNMs, in the limit of
highly damped modes (i.e., QNMs with a large imaginary
part), tends to a constant, that is in agreement with
Refs.\cite{Nomura:2005af} and \cite{Daghigh:2005ph}. This result
satisfies the Hod's conjecture \cite{Hod:1998vk}.

\section{\label{sec:Dilatonic2}Dilatonic Black Hole in Five Dimensions}
There is a growing interest in five-dimensional dilatonic black holes
in last years, since it is
believed that these black holes could shed some light on the
fundamental problem of the microscopic origin of the
Bekenstein-Hawking entropy. The area-entropy relation $S_{BH} = A/4$
was obtained for a class of five-dimensional extremal black holes in
Type II string theory, using D-brane techniques \cite{Strominger:1996sh}.
Also, in
Ref. \cite{Teo:1998kp}, the U-duality that exists between
the five-dimensional black hole and the two-dimensional charged
black hole \cite{McGuigan:1991qp} was used to microscopically compute the
entropy of the latter.

The metric of the five-dimensional
dilatonic black hole can be written as \cite{Teo:1998kp}
\begin{equation}
ds^2=-\left(1-{r_0^2\over r^2}\right)
\left(1+{r_0^2\sinh^2\alpha\over r^2}\right)^{-2}{\rm d}t^2+\left(
{r^2\over r_0^2}-1\right)^{-1}{\rm d}r^2+r_0^2{\rm
d}\Omega_3{}^2.\label{5metric}
\end{equation}

this metric is the product of the two completely decoupled parts,
namely, an asymptotically flat two-dimensional geometry which
describes a two-dimensional charged dilatonic black hole and a
three-sphere with constant radius. This statement can be directly
show  if we apply in the $(t,r)$ sector the transformation defined
by
\begin{equation}
e^{\frac{2}{r_{0}}x}=2\left( \frac{r^{2}}{r_{0}^{2}}+\sinh
^{2}\alpha \right) (m^{2}-q^{2})^{1/2},\label{a1}
\end{equation}
where $m$ and $q$ are related to the mass and charge of the
dilatonic black hole \cite{McGuigan:1991qp}, then Eq.
(\ref{5metric}) read as follow
\begin{equation}
ds^{2}=-N^{2}dt^{2}+N^{-2}dx^{2}+r_{0}^{2}\mathrm{d}\Omega
_{3}{}^{2},\label{2-metric},
\end{equation}
with
\begin{equation}
N^{2}=1-2me^{-Qx}+q^{2}e^{-2Qx}. \label{shift}
\end{equation} Now
we consider the uncharged dilatonic black hole metric, with $q=0$,
\begin{equation}
ds^{2}=-(1-2me^{-Qx})dt^{2}+\frac{dx^{2}}{1-2me^{-Qx}},
\label{2metric}
\end{equation}
as the two dimensional sector of five dimensional dilatonic black
hole that we are interested to compute the QNM's. For complete
this issue we need to solve the equation of motion associated to
the action\begin{equation} S[\varphi ]=-\frac{1}{2}\int
d^{5}x\sqrt{-g}\left( (\nabla \varphi )^{2}+\left( m^{2}+\zeta
R\right) \varphi ^{2}\right) , \label{accion3}
\end{equation}
where $\zeta $ is  a parameter from non-minimal coupling. The
field equation reads as follows,
\begin{equation}
\left( \Box -\mu ^{2}-\zeta'R+\nabla^{2}_{(S^3)}\right) \varphi
=0, \label{fieldequation}
\end{equation}
where $\mu =r_{0}m$, $\zeta'=\frac{\zeta}{r_{0}}$ and
$\nabla^{2}_{(S^3)}$ is the Laplace-Beltrami operator in the $S^3$
sphere. We adopt the following ansatz,
\begin{equation}
\varphi \sim \Phi (t,x)Y(\chi, \theta, \phi), \label{anzatz2}
\end{equation}
where  $Y$ is a normalizable harmonic function on $S^{3}$, i.e., it
satisfies the equation $\nabla^{2}_{(S^3)}Y=\alpha Y$, that in terms of the
coordinates in $S^{3}$ can be written as
\begin{equation}
ccs^{2}\chi \left( \frac{\partial }{\partial \chi }\left( \sin
^{2}\chi
\frac{\partial Y}{\partial \chi }\right) +ccs^{2}\theta \left( \frac{%
\partial }{\partial \theta }\left( \sin ^{2}\theta \frac{\partial Y}{%
\partial \theta }\right) \right) +ccs\theta \frac{\partial ^{2}Y}{\partial
\phi ^{2}}\right) =\alpha Y^{(nlm)},  \label{eigeneq}
\end{equation}
and its solutions are given by
\begin{equation}
Y^{(nlm)}\left( \chi ,\theta ,\phi \right) =\left( \frac{%
2^{2l+1}(n+1)(n-l)!l!^{2}}{\pi (n+l+1)!}\right) \sin ^{l}\chi
C_{n-l}^{(l+1)}\left( \cos \chi \right) Y^{(lm)}(\theta ,\phi ).
\label{egnefunciton}
\end{equation}
Here, $C_{n-l}^{(l+1)}\left( \cos \chi \right) $ are the
Gegenbauer polynomials \cite{Abra,vernon},
$Y^{(lm)}(\theta ,\phi )$ are the $S^{2}$ scalar harmonics, and
the coefficient is chosen to normalize the harmonics. The
eigenvalues are
\begin{equation}
\alpha =-n(n+2),\text{ }\left| m\right| \leq l\leq n=0,1,2,....
\label{eigenvalues}
\end{equation}
Therefore, in this ansatz, we can write Eq.(\ref{fieldequation}) in the following
form,
\begin{equation}
\left( \Box -\mu ^{2}-\zeta'R+n(n+2)\right)\Phi (t,x)  =0,
\label{fieldequation2}
\end{equation}
that is identical to Eq.(\ref{fieldequations}) where the
term $n(n+2)$ is an additive constant. If we repeat the analysis
made in the previous section, we find that the frequencies of the
QNMs are given by
\begin{equation}
\omega_{5D}  =-\frac{i}{4}\left( 1-\sqrt{1-4\zeta ^{\prime
}}-\frac{(1+\sqrt{1-4\zeta ^{\prime }})\mu
^{2}-n(n+2)}{n'+n'^{2}+\zeta ^{\prime }} + n' \left(2-\frac{2\mu
^{2}-2n(n+2)}{n'+n'^{2}+\zeta ^{\prime}}\right)
\right),\label{frecuencias5d}
\end{equation}
with $n$ and $n'$ integer numbers.
\begin{figure}[th]
\includegraphics[width=5.0in,height=5.0in,angle=0,clip=true]{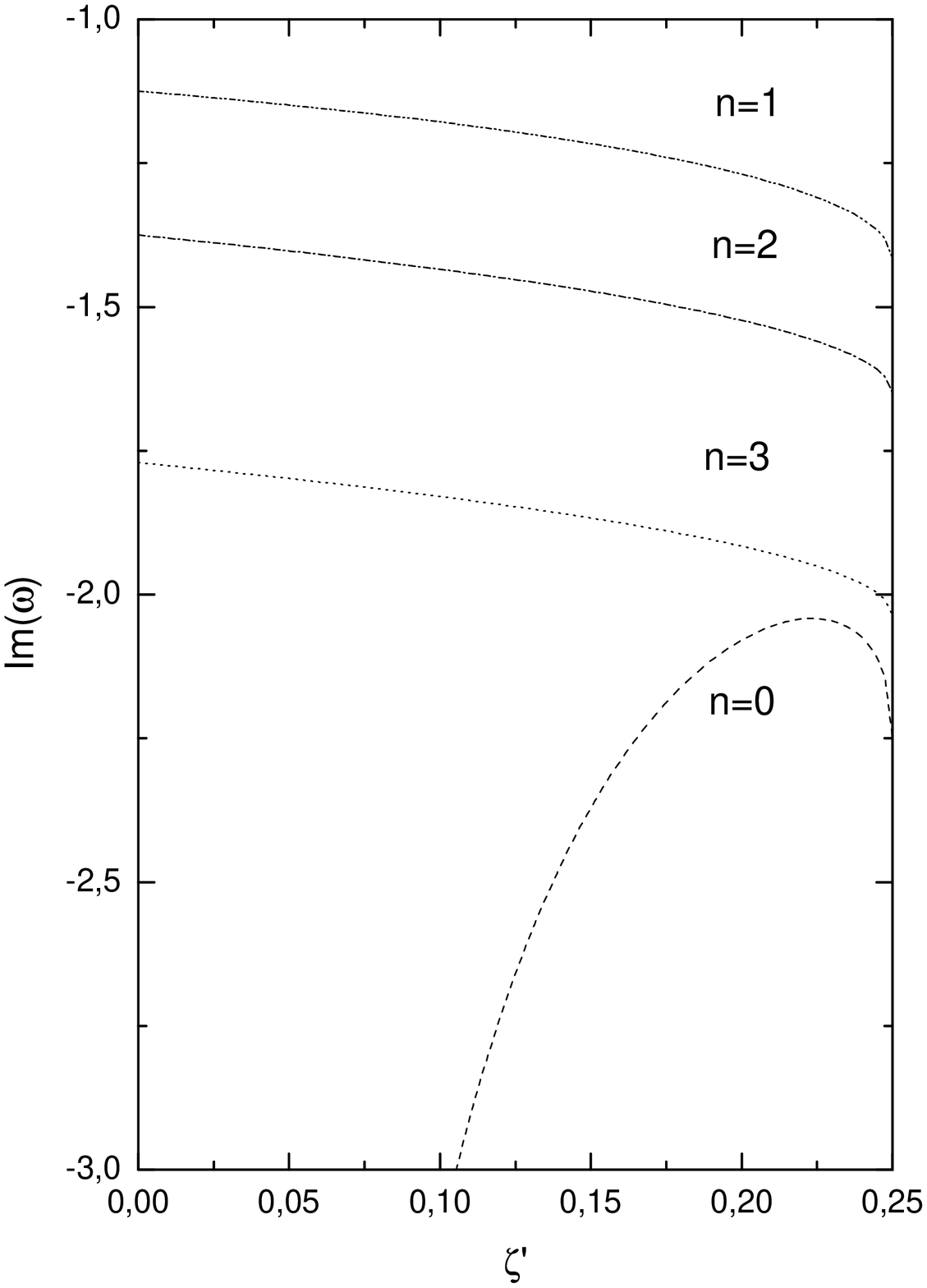}
\caption{The imaginary part of the QNM's frequency as a function
of the no-minimal coupling parameter is illustrated for several overtones. This
plot shows the stable behavior of scalar perturbations for all
overtones of the five-dimensional dilatonic black hole.}
\label{fig4}
\end{figure}
The last expression shows a behavior similar to the one of
the two-dimensional black hole in the range $0\leq
\zeta'\leq1/4$, when $n=0$. If
$n\neq 0$, the situation is completely different due to the
inclusion of transverse part that ensures the stability of the
five-dimensional black hole over all QNMs. This result is shown in
Fig.(4) for $n=1$ and $\mu=1$.

In the range $\zeta'>1/4$, a behavior similar to the one of two-dimensional
case is obtained, that is, the QNMs acquire the same real and
imaginary parts, and the inclusion of the transverse term ensures
the stability in this case, as well. Note that, in the limit of high damping,
the real part tends to the same constant as in the
two-dimensional case.

\section{\label{sec:Dilatonic2}Final Remarks}

In this paper we computed the exact values of the quasinormal
modes of dilatonic black holes in $1+1$ and $4+1$ dimensions and
we showed that the QNMs are purely imaginary (this kind of QNMs was
also reported in Refs.\cite{Lopez-Ortega:2006my,Lopez-Ortega:2006ig,Saavedra:2005ug,
Lopez-Ortega:2005ep,Berti:2003ud,Fernando:2003ai}) in
the range $0\leq \zeta'\leq1/4$ for the no-minimal coupling
parameter. For values of this parameter in the range $\zeta' >1/4$,
we found that the QNMs acquire real parts in both two- and five-dimensional
cases, and in the limit of higher damping they tend to the same constant.
This result is in agreement with the Hod's conjecture
\cite{Hod:1998vk} and it also matches with the results obtained in
Ref.\cite{Nomura:2005af} using the WKB approximation, and in
Ref.\cite{Daghigh:2005ph} where the  monodromy approach was adopted.
Since the considerd kind of black hole does not exhibit a real part in QNMs
in the
range $0\leq \zeta'\leq1/4$, it means that a verification of
the Hod's proposal depends on the values of the no-minimal coupling
parameter. Thus, the Hod's conjecture is no clear at
present, and is fully applicable for a single horizon black hole
obtained in pure Einstein gravity theory. Besides, we found that this
geometry is unstable under scalar perturbations that excite the
zero modes. On the other hand, the result shows the large
stabilities of the dilatonic black hole for the perturbations that
excite the overtones with $n>\mu $ in the minimal case, and
$n>\mu -1/2$ in the conformal case, where all
overtones are taken in the higher damping limit.
Finally, we would like to emphasize that
this result can also be applied to compute the QNMs in
five-dimensional case \cite{Teo:1998kp,McGuigan:1991qp}, where
the metric is the product of a two-dimensional asymptotically flat
geometry and a three-sphere with constant
radius, where these two parts are completely decoupled from each
other.

\begin{acknowledgments}
We are grateful to G. Giribet and O. Miskovic for many useful
discussions, helpful comments and criticism. The authors
acknowledge the referee for useful suggestions in order to improve
the presentation of the results of this paper. This work was
supported by COMISION NACIONAL DE CIENCIAS Y TECNOLOGIA through
FONDECYT \ Grant 1040229 (SL) and 11060515 (JS). This work was
also partially supported by PUCV Grant No. 123.784/2006 (SL) and
No. 123.785/2005 (JS); and by Proyecto SEMILLA UdeC-PUCV Nºs
123.105 and 123.106. R. B. was supported by CONICYT Scholarship
2006.
\end{acknowledgments}

\end{document}